\documentclass[aps,prb,twocolumn,superscriptaddress,10pt]{revtex4-1}
\setcitestyle{numbers,square}
\pdfoutput=1
\usepackage{amsmath,amssymb,amsfonts,upgreek}
\usepackage{color}
\usepackage{graphicx}

\usepackage{prettyref}%
	\newcommand{\pref}[1]{\prettyref{#1}}%
	\newrefformat{fig}{Fig.~\ref{#1}}%
	\newrefformat{tab}{Table~\ref{#1}}%
	\newrefformat{sec}{Sec.~\ref{#1}}%
	\newrefformat{eq}{Eq.~(\ref{#1})}%

\newcommand{\lvo}{LaVO$_3$}
\newcommand{\lto}{LaTiO$_3$}

\graphicspath{{./img/}}

\begin{document}

\title{Charge transfer in LaVO$_3$/LaTiO$_3$ multilayers: strain-controlled dimensionality of interface metallicity between two Mott insulators}

\author{Sophie Beck}
\email{sophie.beck@mat.ethz.ch}
\affiliation{Materials Theory, ETH Z\"u{}rich, Wolfgang-Pauli-Strasse 27, 8093 Z\"u{}rich, Switzerland}
\author{Claude Ederer}
\email{claude.ederer@mat.ethz.ch}
\affiliation{Materials Theory, ETH Z\"u{}rich, Wolfgang-Pauli-Strasse 27, 8093 Z\"u{}rich, Switzerland}

\date{\today}

\begin{abstract}
We use density functional theory plus dynamical mean-field theory to demonstrate the emergence of a metallic layer at the interface between the two Mott insulators LaTiO$_3$ and LaVO$_3$. The metallic layer is due to charge transfer across the interface, which alters the valence state of the transition metal cations close to the interface. Somewhat counter-intuitively, the charge is transferred from the Ti cations with formal $d^1$ electron configuration to the the V cations with formal $d^2$ configuration, thereby increasing the occupation difference of the $t_{2g}$ states. This can be understood as a result of a gradual transition of the charge transfer energy, or electronegativity, across the interface. The spatial extension of the metallic layer, in particular towards the LaTiO$_3$ side, can be controlled by epitaxial strain, with tensile strain leading to a localization within a thickness of only two unit cells. Our results open up a new route for creating a tunable quasi-two-dimensional electron gas in materials with strong electronic correlations.
\end{abstract}

\maketitle

\section{Introduction}

Complex oxide heterostructures have attracted much attention during recent years, in particular due to new functionalities emerging at their interfaces, which are often non-existent in the corresponding bulk components~\cite{Mannhart/Schlom:2010,Hwang_et_al:2012}.
Perhaps the most prominent example of such behavior is the quasi-two-dimensional electron gas (2DEG) emerging at the interface between the two bulk insulators LaAlO$_3$ and SrTiO$_3$~\cite{ohtomo2004high}. 
The emergence of this spatially confined metallic layer can be controlled by an applied electric field~\cite{Thiel_et_al:2006}, and can exhibit magnetic properties~\cite{Brinkman_et_al:2007} as well as superconductivity~\cite{Reyren_et_al:2007}. Interface metallicity has also been reported for other perovskite heterostructures, such as, e.g, SrTiO$_3$/LaTiO$_3$ superlattices~\cite{ohtomo2002artificial}. 
While LaAlO$_3$ and SrTiO$_3$ are both classical band insulators, \lto{} is a Mott insulator, which is insulating due to strong electronic correlations, in spite of its partially filled $d$ band. This raises questions regarding the role of strong interactions for the emergence of the 2DEG, and of possible fundamental differences of interfaces between different types of insulators.

Here, we predict the emergence of a metallic layer at the interface between two Mott insulators, \lto{} and \lvo{}, and analyze the underlying mechanism within the framework of density functional theory combined with dynamical mean-field theory (DFT+DMFT)~\cite{Georges_et_al:1996,Anisimov_et_al:1997,Held:2007}.
\lto{} and \lvo{} are prototypical Mott insulators with $d^1$ and $d^2$ electron configurations of the transition metal (TM) cations, respectively.
Both materials exhibit a perovskite crystal structure with similar magnitude of the so-called GdFeO$_3$-type distortion, manifesting in tilts and rotations of the anion octahedral network, and reducing the space group symmetry to $Pbnm$. In addition, there is only a moderate lattice mismatch of $\sim 1\,\%$ between the two materials, allowing for films with near bulk-like geometries and limiting the effect of strain as an influencing factor. Furthermore, the absence of a polar discontinuity at the \lto{}/\lvo{} interface excludes ambiguities due to specific electrostatic boundary conditions imposed by the periodicity of the used supercells, and puts the focus on the change of the $d$ level occupation of the TM cations across the interface as the main driving force behind the observed behavior. 

Our DFT calculations indicate a charge transfer across the interface from the less occupied Ti cation to the more occupied V cation, consistent with recent suggestions that the O-$p$ states tend to align across the interface~\cite{zhong2017band,Chen/Millis:2017}. As a result, charge transfer is expected from the material with the lower lying  O-$p$ levels to the one with the higher lying ones.
Our analysis of the \lvo{}/\lto{} multilayers confirms this suggestion, but also points towards a smooth and continuous evolution of the charge transfer energy as another important factor.

Our DFT+DMFT calculations then demonstrate that the charge transfer, and the resulting deviation from integer $d$ level occupation, destroys the paramagnetic Mott insulating state in the interfacial layers and leads to a 2DEG at the interface. The localization of this metallic layer in the direction perpendicular to the interface can be controlled through the in-plane epitaxial constraint, with tensile strain leading to a localization within essentially two TMO$_2$ layers, while compressive strain allows the metallicity to penetrate deep into the \lto{} film.
Thus, our study suggests a highly tunable metallic layer, which, due to its embedding between two correlated Mott insulators, might be particularly prone to exhibit unconventional magnetic and transport properties or other effects of strong correlations, paving the way towards future ``Mottronic'' devices.

The remainder of this article is structured as follows.
We first introduce in Sec.~\ref{sec:cm} the theoretical and computational framework used in this study.
In Sec.~\ref{sec:dft}, we then discuss the structural and electronic modifications emerging from the multilayer geometry at the DFT level, and we analyze in particular the band alignment and resulting charge transfer across the interface.
The electronic properties obtained from the consecutive DMFT calculations are then presented in Sec.~\ref{sec:dmft}, and finally Sec.~\ref{sec:sum} summarizes our main results and conclusions.

\section{Computational method}
\label{sec:cm}

To address the interface properties, we use symmetric supercell geometries corresponding to periodic (\lvo{})$_i$/(\lto{})$_j$ multilayers, where $i$ and $j$ are the numbers of \lvo{} and \lto{} perovskite layers, respectively (see Fig.~\ref{fig:ang}a for $i=j=2$).
We use $\sqrt{2} \times \sqrt{2}$ in plane lattice vectors relative to the simple perosvkite units, to allow for the octahedral tilt pattern (a$^-$a$^-$c$^+$) present in both bulk materials, with the layers stacked along the long orthorhombic axis of the bulk $Pbnm$ unit cell, which defines the $c$-axis of the slab.
For even numbers $i$ and $j$, this geometry preserves both the glide plane $b$ parallel to $c$, and the mirror plane $m$ perpendicular to $c$ within the central LaO layers in each material, resulting in $(i+j)/2$ symmetry-inequivalent TM sites within the supercell. We note that an even number of $i+j$ layers does not interfere with the bulk tilt pattern, and thereby allows for bulk-like tilts throughout the whole slab.
We use the notation ``$i$/$j$'' in the following to denote slabs with different thickness of the individual layers.

We consider different cases, where we fix the in-plane cell parameter to specific values, while the $c$-component of the cell and all internal coordinates are fully relaxed within DFT.
In the ``unstrained'' case, we fix both in-plane cell parameters according to the average of the calculated pseudocubic lattice constants, \mbox{$a_{\text{cub}} = (a+b+c/\sqrt{2})/3\sqrt{2}$}, of the two bulk materials.
Our calculated pseudocubic bulk lattice constants for \lvo{} and \lto{}, $a_{\text{LVO}}=3.894$\,\AA{} and $a_{\text{LTO}}=3.960$\,\AA, result in an average in-plane cell parameter, $a_{\text{avr}}=3.927$\,\AA, which corresponds to an effective tensile (compressive) epitaxial strain of $+0.8$\,\% ($-0.8$\,\%) for the \lvo{} (\lto{}) layers.
To investigate the effect of different substrate-induced strains, we also consider in-plane lattice parameters that are larger or smaller than $a_\text{avr}$.
Here, we use as reference two common substrate materials, KTaO$_3$ and LaAlO$_3$, with calculated lattice parameters larger, respectively smaller than $a_\text{avr}$.
The corresponding lattice constants and the resulting strains in the \lvo{} and \lto{} layers for the three different cases are summarized in Table~\ref{tab:strain}.
In addition we compare the relaxed system at the averaged in-plane cell parameter to one with suppressed octahedral rotations.
This way we aim to address the effect of the octahedral distortions on the length scale and magnitude of the interfacial charge transfer.
\begin{table}[t]
  \caption{Different in-plane lattice parameters and resulting effective epitaxial strains $\eta_\text{LVO}$ and $\eta_\text{LTO}$ in the \lvo{} and \lto{} layers, respectively, relative to their pseudocubic bulk lattice constants. The value of $a_{\text{avr}}$ is the calculated average pseudocubic lattice constant of both compounds, while $a_\text{KTO}$ and $a_\text{LAO}$ refer to KTaO$_3$, and LaAlO$_3$, respectively.}
  \label{tab:strain}
  \setlength{\tabcolsep}{5pt}
  \begin{tabular}{lccc}
    \hline\hline
     & $a_{\text{avr}}$ & $a_{\text{KTO}}$ & $a_{\text{LAO}}$  \\ \hline
    $a$ [\AA] & 3.927 & 4.027 & 3.812   \\ 
    $\eta_{\text{LVO}}$ [\%] & 0.8 & 3.4 & -2.1   \\ 
    $\eta_{\text{LTO}}$ [\%] & -0.8 & 1.7 & -3.7 \\ \hline\hline
  \end{tabular}
\end{table}

All DFT calculations are performed using the plane-wave based \textsc{Quantum~ESPRESSO} package~\cite{Giannozzi_et_al:2009}, scalar-relativistic ultrasoft pseudopotentials, and the generalized gradient approximation for the exchange-correlation functional as parametrized by Perdew, Burke, and Ernzerhof (PBE)~\cite{Perdew/Burke/Ernzerhof:1996}. 
Plane waves are included up to a kinetic energy cutoff of 70 Ry for the wavefunctions, and 840 Ry for the charge density.
A $6 \times 6 \times 3$ Monkhorst-Pack $k$-point grid is used in all cases except for the 1/1 slab, for which we use a $6 \times 6 \times 4$ grid.
Brillouin-zone integrals are evaluated with a Methfessel-Paxton smearing parameter of 0.02\,Ry for atomic relaxations, and 0.01\,Ry for all other calculations.
The 3s and 3p semicore states of V and Ti, and the 5s and 5p semicore states of La, are included in the valence.
Atomic positions are relaxed until all force components are smaller than 0.1 mRy/$a_0$ ($a_0$: Bohr radius), and the cell parameter along $c$ is adjusted until the corresponding components of the stress tensor are smaller than 0.1 kbar.

All DFT calcuations are performed without considering spin polarization, leading to metallic Kohn-Sham bandstructures.
We then use the \textsc{Wannier90} code \cite{Mostofi_et_al:2008} to construct maximally localized Wannier functions (MLWFs)~\cite{Marzari_et_al:2012} describing the partially filled TM-$t_{2g}$-dominated bands around the Fermi level.
The resulting low-energy tight-binding Hamiltonian, corresponding to three $t_{2g}$-like MLWFs per TM site, serves as input for the consecutive DMFT calculations, where we incorporate the effect of local electronic correlations.
Each inequivalent TM site is mapped on an effective local impurity problem, which is solved using the TRIQS/CTHYB solver~\cite{Seth2016274}, while different impurity problems are connected self-consistently via the lattice Green's function.
The local electron-electron interaction is described by the Slater-Kanamori interaction Hamiltonian, including spin-flip and pair-hopping terms, and using a value of $J=0.65$\,eV for the Hund's coupling~\cite{Sclauzero/Dymkowski/Ederer:2016}.
The value of the Hubbard $U$ is varied systematically to elucidate the resulting change in the physical properties. For simplicity we use the same $U$ for both materials and all layers.
We note that the $U$ values which have been used in previous DFT+DMFT studies were similar for both \lvo{} and \lto{}~\cite{Pavarini_et_al:2004,DeRaychaudhury/Pavarini/Andersen:2007,Dymkowski/Ederer:2014,Sclauzero/Ederer:2015}.
Self-consistency is implemented using the TRIQS/DFTTools libraries~\cite{aichhorn_dfttools_2016}, yielding the local impurity Green's functions in imaginary time, $G(\tau)$, from which we deduce the orbital occupations, \mbox{$n = -G(\beta)$}, where $\beta=1/(k_{\text{B}}T)$ is the inverse temperature, set to $40$ eV$^{-1}$.
Local spectral features are evaluated from the ``averaged'' spectral density at the Fermi level, \mbox{$\bar{A}(0) = - \tfrac{\beta}{\pi} \textrm{Tr}[G(\beta/2)]$}, which approaches the spectral function, $A(\omega = 0)$, for decreasing temperature~\cite{Fuchs_et_al:2011}, or by calculating the full real-frequency spectral function, $A(\omega)$, via analytic continuation using the maximum entropy method~\cite{jarrell1996bayesian}.
To reduce the computational cost we limit the calculations to a single-shot DMFT self-consistency loop, without updating the DFT band structure, i.e., neglecting full charge self-consistency.

\section{DFT results --- Multilayer structure and charge transfer}
\label{sec:dft}
\subsection{Structural properties}

\begin{figure}
    \centering
    \includegraphics[width=\linewidth]{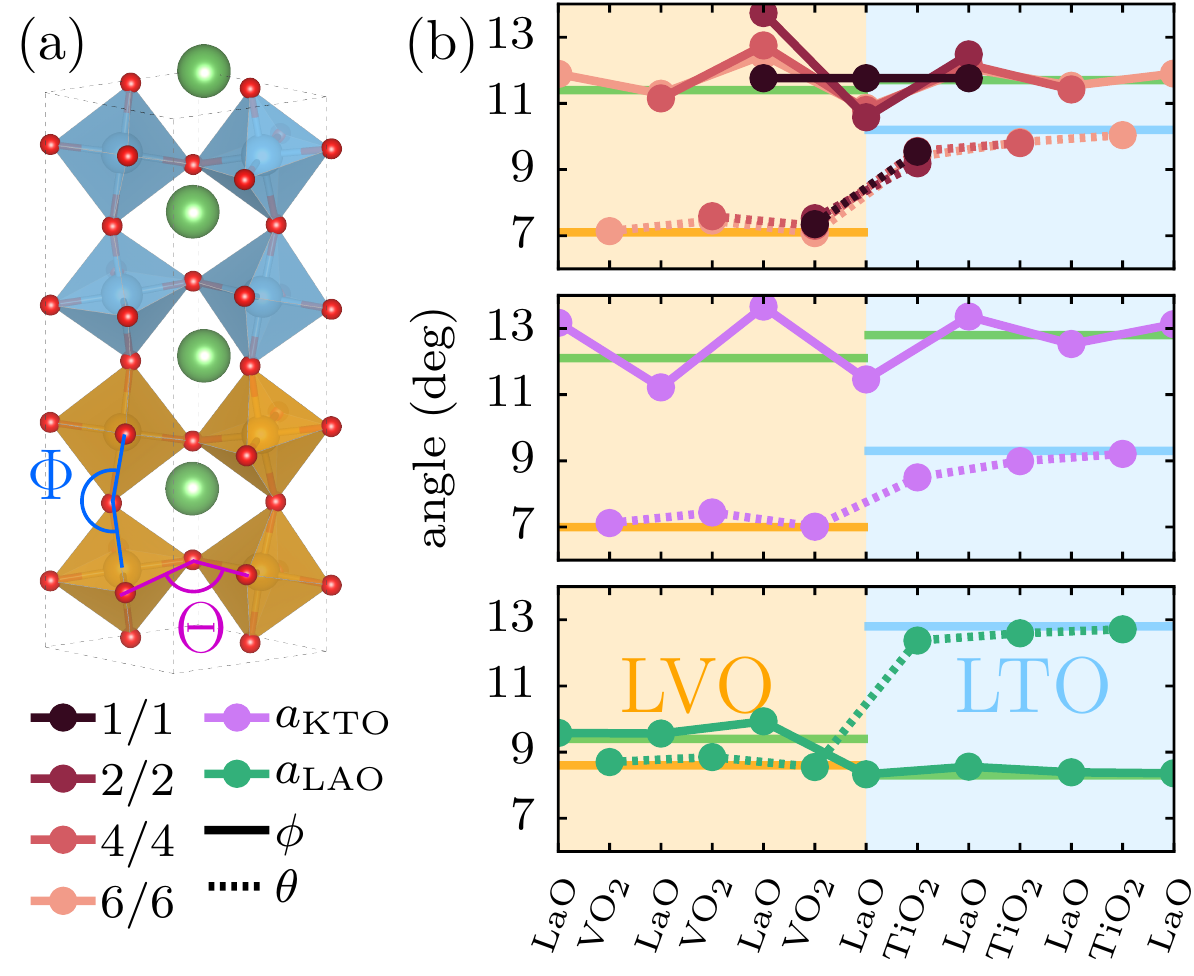}%
    \caption{%
    (a) Unit cell of the 2/2 multilayer. La atoms are shown in green, V (Ti) in orange (blue), and O in red. The TM-O-TM and O-O-O bond angles $\Phi$ and $\Theta$ used to define the octahedral tilt and rotation angles $\phi=(\pi-\Phi)/2$ and $\theta=(\pi/2-\Theta)/2$ are illustrated.
    (b) Evolution of octahedral tilts and rotations in the \lvo{}-\lto{} multilayers along the $c$-direction, represented by solid and dashed lines, respectively. From top to bottom panel: multilayers with the in-plane cell parameters set to $a_{\text{avr}}$, $a_{\text{KTO}}$ (tensile strain), and $a_{\text{LAO}}$ (compressive strain). The horizontal lines indicate the corresponding bulk values.}
    \label{fig:ang}
\end{figure}

We first analyze the structural modifications in our DFT-relaxed supercells, focusing on the effect of the layering on the octahedral tilt patterns of both compounds in terms of layer-resolved tilt ($a^-a^-c^0$) and rotation ($a^0a^0c^+$) angles, $\phi$ and $\theta$, as indicated in Fig.~\ref{fig:ang}a.
The former is typically defined from the TM--O--TM bond angle along $c$, while the latter is related to an in-plane O--O--O angle between two corner-sharing octahedra, as described, e.g., in Ref.~\onlinecite{Rondinelli:2011}.
Fig.~\ref{fig:ang}b shows the evolution of these angles with the distance from the interface for different layer thicknesses $i$/$j$, and different in-plane strains.
Also indicated are the corresponding bulk values for both \lvo{} and \lto{}, calculated at the corresponding nominal strains (see Table~\ref{tab:strain}).

It can be seen that the in-plane rotations ($\theta$, dashed lines) adopt the corresponding bulk values on both sides of the interface essentially from one layer to the next, despite differences between these bulk values of up to 4$^\circ$. 
Only in the first TiO$_2$ layer next to the interface, a small reduction compared to the bulk value is noticeable.
This weak effect of the heterointerface on the in-plane rotations can be expected, since, at least in the absence of any tilts, the octahedral rotations around $c$ in one plane do not directly affect the octahedral network in adjacent planes.

The out-of-plane tilt angles ($\phi$, solid lines) coincide within 1$^\circ$ in all three cases, such that also here a smooth transition can be expected. However, with increasing in-plane lattice constant, an oscillatory behavior around the corresponding bulk values can be observed, in particular on the \lvo{} side.
This indicates an additional distortion of the structure, since different TM-O-TM bond angles in adjacent layers are incompatible with completely rigid oxygen octahedra.
However, as discussed further below, these distortions do not seem to be crucial for the electronic interfacial properties, and thus we do not analyze them in more detail.
All in all, it appears that the collective octahedral tilts and rotations in this system are not disturbed much by the presence of the interface.

\subsection{Charge transfer}

\begin{figure}
    \centering
    \includegraphics[width=\linewidth]{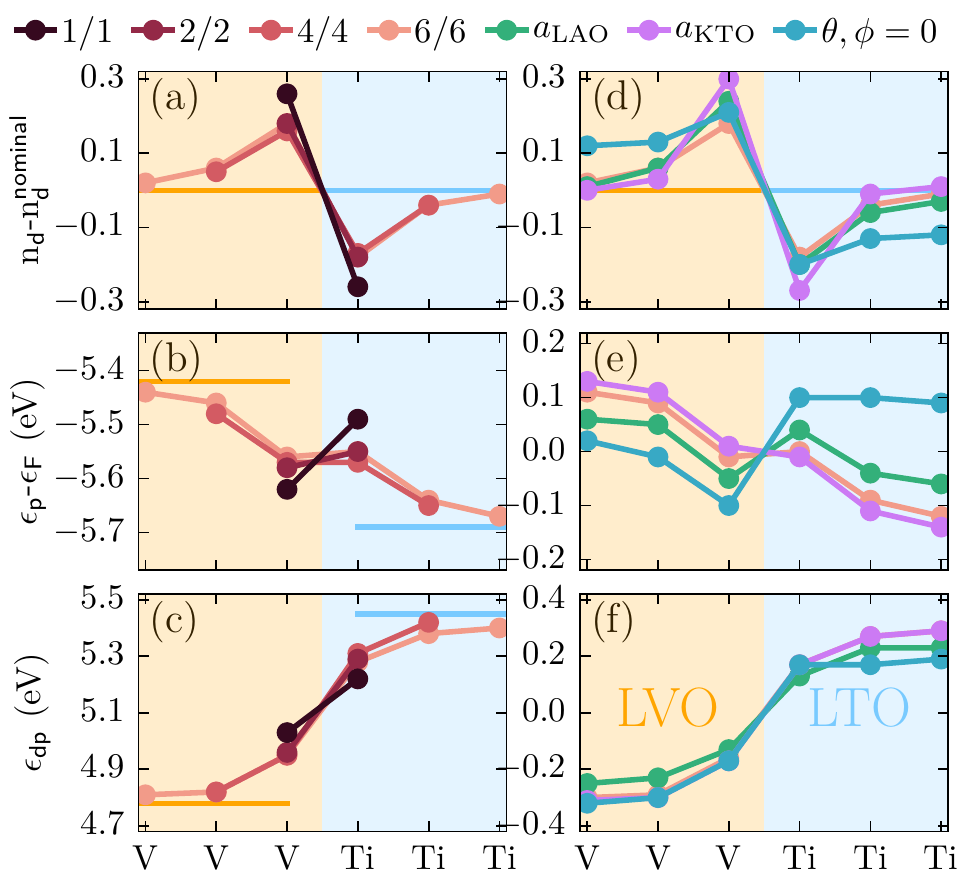}%
    \caption{%
    Evolution of the $d$ occupation changes (top), the O-$p$ energies with respect to the Fermi level (middle) and the energy differences $\epsilon_{dp}$ (bottom) as a function of the TM site along the $c$-direction of the multilayers.
    The left panels show multilayers of various thickness at $a_{\text{avr}}$, the right side only concerns the largest 6/6 heterostructures, comparing different strains corresponding to $a_{\text{avr}}$, $a_{\text{KTO}}$ (tensile), and $a_{\text{LAO}}$ (compressive), as well as the case without octahedral rotations ($\theta,\phi=0$).
    To allow for a better comparison, the curves in panels (e-f) are centered around the respective mean values of the interface layers.
    }
    \label{fig:dft}
\end{figure}

Next, we turn to the charge transfer at the interface by analyzing the electronic structure obtained within plain DFT, which serves as input for our DMFT calculations discussed in Sec.~\ref{sec:dmft}.
Fig.~\ref{fig:dft}a shows the layer-dependent change in occupation with respect to the formal $d$ electron occupations of 2 and 1 for \lvo{} and \lto{}, respectively.
These occupations correspond to the $t_{2g}$-like MLWFs constructed for the partially filled bands around the Fermi level.
For the bulk systems these occupations are identical to the formal $d$ occupations.

Naively, one might assume to find a monotonous transition of the $d$-level occupations at a $d^2$/$d^1$ interface, resulting in a mixed occupational state $d^x$ with $1<x<2$ of the interfacial layers, as was demonstrated, e.g., in SrVO$_3$/\lvo{} superlattices, both theoretically~\cite{park2017charge} and experimentally~\cite{tan2013mapping}.
However, our calculations show that for all considered multilayers the charge transfer at the \lvo{}/\lto{} interface is in the opposite direction, i.e., charge is transferred from the Ti cations in \lto{} to the V cations in \lvo{}, thereby increasing the charge imbalance of the $d$ states at the interface. 
This is also in qualitative agreement with recent DFT+$U$ calculations for magnetically ordered \lvo{}/\lto{} superlattices~\cite{weng2017latio}.
Furthermore, the charge transfer is strongly localized at the interface, converging back to bulk-like occupations within only three layers.

\begin{figure}
    \centering
    \includegraphics[width=.6\linewidth]{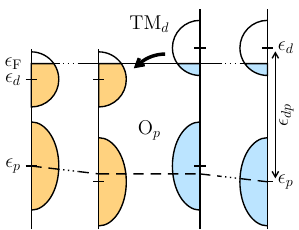}%
    \caption{%
    Schematic figure of the band alignment at the interface.
    The averages of the on-site energies of the O-$p$ and the TM-$d$ bands are given by $\epsilon_p$ and $\epsilon_d$, respectively.
    $\epsilon_{\text{F}}$ indicates the Fermi level.
    Far away from the interface (outermost layers) one expects a bulk-like situation, while at the interface of ultrathin multilayers (inner layers) a smooth transition of $\epsilon_{dp}$ can lead to an inverted alignment of the O-$p$ states.
    }
    \label{fig:ct}
\end{figure}

Different mechanisms for charge transfer at oxide interfaces have been discussed in Refs.~\onlinecite{zhong2017band} and \onlinecite{Chen/Millis:2017}. \citet{zhong2017band} have proposed that the charge transfer can be predicted based on an ``oxygen continuity condition'', which is closely related to the discussion of electronegativity differences by~\citet{Chen/Millis:2017}. 
To further understand the mechanism for the interfacial charge transfer, we therefore construct an additional set of MLWFs, corresponding to an extended energy window including both the correlated low-energy ``TM $t_{2g}$-bands'' as well as as the ``O-$p$'' bands at lower energies. 
The on-site energies of these MLWFs (also illustrated in Fig.~\ref{fig:ct}) have been suggested in Ref.~\onlinecite{zhong2017band} as a reliable measure to analyze the alignment of O-$p$ and TM-$d$ bands across the interface.

Fig.~\ref{fig:dft}b shows the evolution of these energy levels (averaged over all $d$ or $p$ levels in the same TMO$_2$ layer) across the interface in the various multilayer structures, together with the corresponding bulk reference values.
It can be seen that in bulk \lto{} the O-$p$ states lie much lower with respect to the Fermi level than in bulk \lvo{}. 
This can be related to the lower electronegativity of Ti compared to V with respect to the O-$p$ states.
Thus, aligning the O-$p$ levels at the interface, as suggested in Ref.~\onlinecite{zhong2017band}, predicts charge transfer from the Ti$^{3+}$ cations into the $d$ states of the V$^{3+}$ cations, in agreement with the computational results.
Furthermore, the O-$p$ levels in the multilayer structures appear well aligned across the interface, then converge back to the bulk values within a few layers away from the interface, thereby supporting the validity of the oxygen level continuity condition. 

However, the case of the 1/1 superlattice is exceptional, as it exhibits an inverted alignment of the O-$p$ levels relative to the bulk values.
Furthermore, we note that the total amount of transferred charge of 0.26\,$e^-$ is greater or equal compared to those in the larger superlattices with 0.18\,$e^-$ (2/2), 0.21\,$e^-$ (4/4), and 0.26\,$e^-$ (6/6).
These two facts are somewhat counterintuitive if the sign and strength of the charge transfer is mainly determined by the respective bulk O-$p$ energies as proposed in Ref.~\onlinecite{zhong2017band}, and suggests that there are also other mechanisms at play.

To investigate this further, Fig.~\ref{fig:dft}c shows the charge transfer energy, $\epsilon_{dp}=\epsilon_d-\epsilon_p$, which exhibits a very smooth and monotonous evolution between the two bulk values across the interface.
We note that $\epsilon_{dp}$ can also be viewed as a direct measure of the electronegativity of the TM cation (whereby higher $\epsilon_{dp}$ means lower electronegativity).
We further note that the TM-$d$ on-site energies, $\epsilon_d-\epsilon_\text{F}$, are nearly unchanged compared to the bulk values and are almost layer-independent (not shown here), such that $\epsilon_{dp}$ essentially follows $-\epsilon_p$.
The small variation of $\epsilon_d$, despite the transferred charge, is probably related to the high density of states at the Fermi level. 
Considering the smooth variation of the charge transfer energy across the interfacial region (see Fig.~\ref{fig:dft}c), it appears that the inverted alignment of the O-$p$ levels in the 1/1 superlattice allows for a smaller difference of $\epsilon_{dp}$ across the interface, as is depicted schematically in Fig.~\ref{fig:ct}, even if it leads to a stronger deviation from the bulk values.
We note that charge transfer within the $d$ states from the material with larger $\epsilon_{dp}$ to the one with smaller $\epsilon_{dp}$ will tend to equalize the electronegativity of both materials.
That is, the change in electronegativity, $\Delta \epsilon_{dp}$, has the same sign as the transferred charge, $\Delta n$, according to $\Delta \epsilon_{dp} \approx \Delta n U_H$, as proposed in Ref.~\onlinecite{zhong2017band}. Here, $U_H$ represents an effective electron-electron interaction.

Thus, our results suggest that while the alignment of the O-$p$ states (which could be seen as a misalignment in the case of the 1/1 superlattice) might initially act as a driving force for the charge transfer, a smooth and continuous variation of $\epsilon_{dp}$ (indicative of the electronegativity) in the converged state appears to be another relevant factor controlling the band alignment at the interface.
Note that in the larger superlattices, all interfacial TM atoms only share one oxygen ligand bonding with the other type of TM atom, while in the 1/1 superlattice each TM atom shares two oxygen atoms with the other type.
This reduced coordination number of bulk-like oxygen ligands could explain the stronger deviation of $\epsilon_{dp}$ from the respective bulk values.

Next, we analyze the effect of epitaxial strain on the interfacial charge transfer, as well as the case with suppressed octahedral rotations. The corresponding results (only for the 6/6 multilayers) are shown in Fig.~\ref{fig:dft}d-f.
While the layer-resolved $d$ occupations under compressive strain (green) are almost indistinguishably from the unstrained case, tensile strain (purple) increases the charge transfer slightly, while simultaneously the effect appears to be more localized at the interface, such that nominal occupations are recovered already in the first subinterface layer.
With suppressed octahedral rotations, however, we observe the opposite trend, i.e., the transferred charge penetrates deep into the layers, and even for the 6/6 multilayer bulk occupations are not yet recovered in the innermost \lvo{}  and \lto{} layers.
This is likely due to the much larger $t_{2g}$ bandwidth resulting from the ideal 180$^\circ$ TM-O-TM bond angles (about 3.4\,eV compared to 2.5\,eV), which decreases the tendency to locate the charge transfer at the interface.
Fig.~\ref{fig:dft}e also reveals that both under compressive strain and for the case with suppressed octahedral rotations, the O-$p$ levels again show an inverted alignment at the interface, similar to the 1/1 superlattice.
Simultaneously, the charge transfer energy $\epsilon_{dp}$ again exhibits a very smooth and continuous evolution, indicating the potential relevance of an ``$\epsilon_{dp}$ continuity'' across the interface.

\section{DMFT results}
\label{sec:dmft}

\begin{figure}
    \centering
    \includegraphics[width=\linewidth]{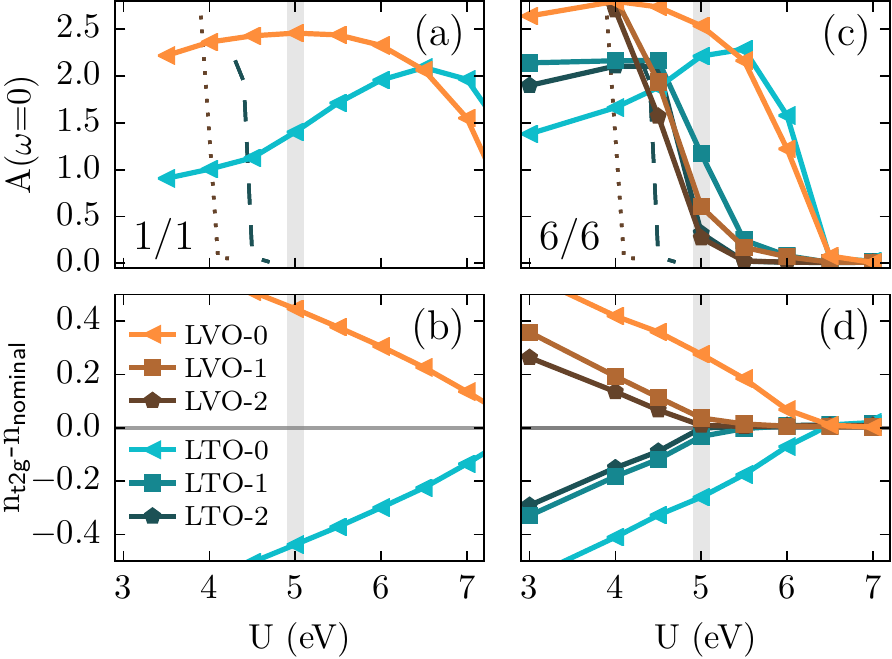}
    \caption{DMFT results for the unstrained 1/1 (left) and 6/6 (right) multilayer.
    (a)+(c) Averaged spectral weight at the Fermi level, $\bar A(0)$, and (b)+(d) total occupation per site as a function of the interaction parameter $U$.
    Orange/brown lines and symbols refer to \lvo{}, while blue/turquoise ones represent \lto{} layers, with $n$ in LVO-$n$ (LTO-$n$) indicating the distance from the interface. The bulk reference data for \lvo{} and \lto{} at $a_{\text{avr}}$ is indicated by dotted and dashed lines, respectively.}
    \label{fig:dmft}
\end{figure}

We now turn to the question of how the charge transfer affects the Mott insulating state of both \lvo{} and \lto{}.
\pref{fig:dmft}a and c shows the layer-resolved spectral weight at the Fermi level, $\bar{A}(0)$, for the unstrained 1/1 and 6/6 multilayers, obtained from our DFT+DMFT calculations as described in Sec.~\ref{sec:cm}.
We use the notation LVO-$n$ (LTO-$n$) with $n \in \{0,1,2\}$ to indicate the proximity to the interface layer on the \lvo{} (\lto{}) side, i.e., to layer LVO-$0$ (LTO-$0$).
Data for the individual bulk systems, with strain-levels corresponding to $a_\text{avr}$ (see Table~\ref{tab:strain}), are indicated by the dotted and dashed lines for \lvo{} and \lto{}, respectively. The bulk systems are found to be metallic for $U < 4.1$\,eV (\lvo{}) and $U < 4.5$\,eV (\lto{}). Thus, for a typical value of $U \approx 5$ eV \cite{Pavarini_et_al:2004, DeRaychaudhury/Pavarini/Andersen:2007, Dymkowski/Ederer:2014, Sclauzero/Ederer:2015}, $\bar{A}(0)$ is zero for both compounds, representing their Mott insulating states.

Considering first the 1/1 superlattice in Fig.~\ref{fig:dmft}a, we find that both layers remain at finite values $\bar A(0)>0$ up to very large $U$ values, suggesting that this superlattice is fully metallic.
This is not surprising since the redistribution of charges leads to non-integer occupations in both layers at the DFT level, thus, making the Mott insulating states less favorable.
Fig.~\ref{fig:dmft}b shows the change of the $d$ occupation in the two layers with varying $U$. It can be seen that increasing $U$ reduces the charge transfer, effectively pushing the $d$ occupations back to the nominal values.
However in the case of the 1/1 multilayer, integer occupations cannot be achieved for realistic $U$ values and thus the systems remains metallic.
One can also see that the deviation from nominal occupations is much stronger than the charge transfer in the DFT calculations shown in Fig.~\ref{fig:dft}.
This is due to the double-counting correction applied to the DFT results before the DMFT calculation, which effectively lowers the energy of the more occupied sites relative to the less occupied sites, and thus enhances the charge transfer.
This effect is then counterbalanced again by the explicit electron-electron interaction among the $d$ electrons.

As shown for the 6/6 multilayer in \pref{fig:dmft}c, increasing the thickness of the \lto{} and \lvo{} layers reduces the value of $U$ for which the spectral weight at the Fermi level vanishes.
However, the degree of this reduction for each layer depends strongly on the distance to the interface.
All layers except the immediate interface layers become insulating at around $U=5$\,eV, whereas both the \lto{} and the \lvo{} interface layers are still metallic up to $U \approx 6$\,eV.
Increasing the Coulomb repulsion further eventually leads to a completely insulating state, however, long after the subinterface layers have undergone the metal-insulator transition (MIT).
Thus, for $U$ values slightly above $5$\,eV, one obtains a metallic interface while the inner layers are already insulating.
These trends are also mirrored by the occupancies shown in \pref{fig:dmft}d, which reach the nominal values in the subinterface layers around $U=5$\,eV, while the interface layers are still subject to the charge transfer.
For thicker multilayers, one can expect that the inner layers will converge further to the insulating bulk state, while the interface layers remain metallic for a realistic $U \approx 5$\,eV.
Thus, our DMFT calculations predict a metallic interface between the two Mott insulators \lvo{} and \lto{}, which emerges from the charge transfer that results in non-integer $d$ occupations close to the interface.

As shown in Refs.~\onlinecite{Dymkowski/Ederer:2014,Sclauzero/Ederer:2015,Sclauzero/Dymkowski/Ederer:2016}, epitaxial strain can have a large effect on the MIT in both \lto{} and \lvo{}.
Therefore, we investigate next how different in-plane constraints affect the layer-dependent MIT in the corresponding heterostructures.
As already shown in Sec.~\ref{sec:dft}, compressive strain has negligible effect on the interfacial charge transfer, while tensile strain leads to a slightly stronger localization of the charge transfer at the interface.

\begin{figure}
    \centering
    \includegraphics[width=\linewidth]{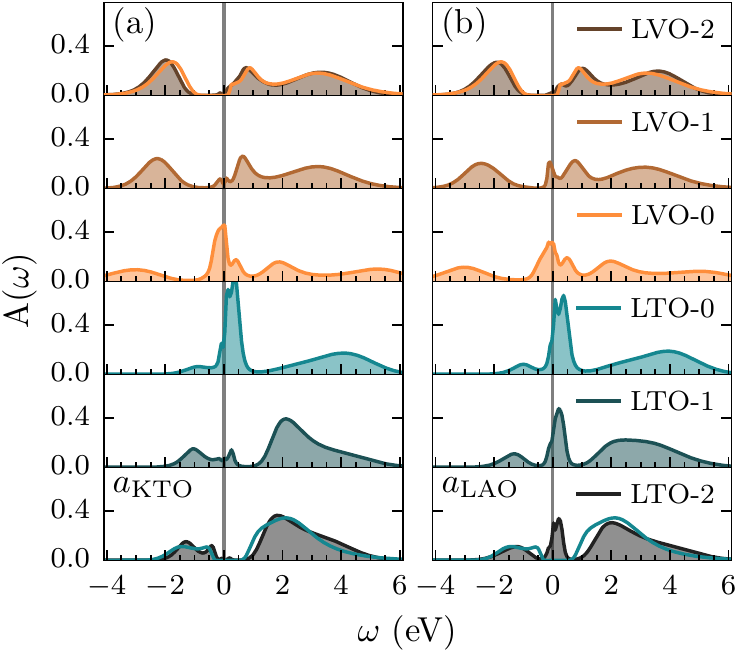}
    \caption{
    Layer-resolved spectral functions of the 6/6 multilayer for tensile (left), and compressive (right) strain at $U=5$ eV. The (unstrained) bulk spectral functions are indicated by bright orange (\lvo{}), and blue (\lto{}) lines in the panels for the innermost layers (LVO-2 and LTO-2).}
    \label{fig:sfunc}
\end{figure}

We first look at the case of tensile strain, with in-plane lattice constant $a_\text{KTO}$ (see \pref{tab:strain}).
According to Ref.~\onlinecite{Sclauzero/Dymkowski/Ederer:2016}, tensile strain will favor the insulating state in both \lvo{} and \lto{}.
This is indeed consistent with the layer-resolved spectral functions, $A(\omega)$, shown in \pref{fig:sfunc}a for $U=5$\,eV.
While the two layers directly at the interface are still metallic, similar to the unstrained case, and exhibit strong quasiparticle features at the Fermi level ($\omega=0$), the metallic character decays very rapidly away from the interface. 
Even the two subinterface layers (LVO-1 and LTO-1) are only very weakly metallic, without strong quasiparticle features.
Finally, the spectral functions of the two innermost layers (LVO-2 and LTO-2) have almost recovered the (unstrained) bulk-like features, indicated by the bright orange and blue lines, resulting in a completely insulating state already two layers away from the interface, both on the \lvo{} and \lto{} sides.

\pref{fig:sfunc}b shows the layer-resolved spectral functions for $U=5$\,eV under compressive strain (in-plane lattice constant $a_\text{LAO}$, see Table~\ref{tab:strain}).
It can be seen that the \lvo{} layers are quite similar compared to the tensile strain case. The interface layer (LVO-0) exhibits clear metallic features, which rapidly decay away from the interface.
That is, while there remains spectral weight at the Fermi level in the first sub-interface layer, LVO-2 has already recovered Mott-insulating bulk behaviour.
However, on the \lto{} side, the quasiparticle features remain visible in all three layers.
This indicates that, while in \lvo{} the metallic state is limited to the interface, the strong compressive strain has lead to an overall metallic \lto{} film, consistent with the MIT in \lto{} thin films under compressive strain predicted theoretically~\cite{Dymkowski/Ederer:2014} and reported experimentally~\cite{He_et_al:2012}.
It further demonstrates, that the dimensionality of the metallic layers in \lvo{}/\lto{} heterostructures can be tuned through epitaxial strain, reaching from a thickness of essentially two unit cells for tensile, to the thickness of the entire \lto{} film under compressive strain.

\section{Summary and conclusion}
\label{sec:sum}

In conclusion, we report on a metallic interface between two prototypical Mott insulators, \lvo{} and \lto{}, whose dimensionality can be adjusted by epitaxial strain.
The metallic interface results from the charge transfer across the interface from \lto{} towards \lvo{}, which can be understood in terms of their electronegativity difference related to  $\epsilon_{dp}$.
We demonstrate that in most cases the band alignment is in line with the oxygen continuity condition proposed in Ref.~\onlinecite{zhong2017band}, i.e., that the O-$p$ levels tend to align across the interface, and that their bulk values correctly predict the direction of the charge transfer.
However, our analysis also reveals that another instructive indicator for the charge transfer is given by the energy difference $\epsilon_{dp}$ between the O-$p$ and the TM-$d$ states, showing a smooth transition across the interface for all studied multilayers, even when the oxygen continuity condition is no longer obvious as, e.g., in the case of the 1/1 superlattice.

On a structural level, we showed that the octahedral tilts and rotations converge to their bulk values within only two layers (except, perhaps, for the case of tensile strain, where we observed some deviations from a rigid octahedral tilt distortion), i.e., the octahedral rotations in \lvo/\lto{} multilayers are only weakly affected by the interface.
Nevertheless, heterostructuring can have a strong influence in the case of less compatible tilt systems.
For example, it was shown experimentally that octahedral rotations in \lvo{} can be suppressed in multilayers upon insertion of a critical number of layers of cubic SrVO$_3$~\cite{Luders_et_al:2014}. 
Although in the present case the suppression of octahedral rotations appears somewhat artifical, we found that it would render the entire superlattice metallic due to the strong increase of the bandwidth upon straightening out the TM-O-TM bond angles.

We are not aware of any systematic experimental studies of charge transfer and its effect on the Mott insulating behavior at the \lvo{}/\lto{} interface.
A study of LaTi$_{1-x}$V$_x$O$_3$ solid solutions found that all compositions remained insulating, except in the range of $0.1<x<0.25$, for which the samples showed (poor) metallic behaviour~\cite{eylem1996unusual}. This was interpreted as a result of the variation of the bandwidth in competition with a non-linear change in the Coulomb repulsion.
Considering that we find the 1/1 superlattice to be metallic, a more systematic study comparing the properties of bulk solid solutions with short-period superlattices of similar composition would be instructive.
Note that DFT+$U$ calculations for magnetically ordered \lvo{}/\lto{} superlattices also found the 1/1 case with layering perpendicular to [001] to be metallic, in agreement with our DFT+DMFT results, while other stacking directions turned out to be insulating~\cite{weng2017latio}.
Another study addressed the role of strain and polarity at interfaces of \lvo{} and \lto{} with various substrates~\cite{He_et_al:2012}, and suggested that the insulator-to-metal transition in both materials is due to a complex interplay of structural, and electronic degrees of freedom that affects the two materials in different ways.
However, the transport properties of the \lvo{}/\lto{} interface were not analyzed in detail in that work and thus a comparison with our computational results in not possible.

Our findings presented in this work emphasize the role of internal charge transfer and heterostructuring as control parameters in tailoring emergent phenomena at interfaces.
Reducing the complexity by excluding effects of oxygen vacancies or polar interfaces, we are able to analyze the interplay between purely structural changes, and the electronic reconstruction.
In particular the possibility to adjust the thickness of the metallic layer via epitaxial strain could allow for an efficient tailoring of the resistity and other physical properties.
The Mott insulating character of both components can also lead to other emerging properties such as magnetic order or superconductivity or other effects related to strong electronic correlations.
It was shown in other \lto{}-based multilayers that charge transfer from \lto{} can lead to dramatic changes in the magnetic properties of the second material, for example enhanced correlation effects in LaNiO$_3$~\cite{Chen_et_al:2013}, and even completely suppressed magnetism in LaFeO$_3$~\cite{Kleibeuker_et_al:2014}.
Furthermore, it is conceivable that an applied electric field could be used to (at least partially) reverse the charge transfer, effectively allowing such multilayers to function as field-effect transistors.
These effects open up a multitude of possibilities for future electronic applications and ``Mottronic'' devices that benefit from the strongly correlated nature of these systems.

\appendix*

\begin{acknowledgments}

This work was supported by ETH Zurich and the Swiss National Science Foundation through NCCR-MARVEL. Calculations have been performed on the cluster ``Piz Daint'' hosted by the Swiss National Supercomputing Centre.

\end{acknowledgments}

\bibliography{bibfile}

\begin{thebibliography}{34}%
\makeatletter
\providecommand \@ifxundefined [1]{%
 \@ifx{#1\undefined}
}%
\providecommand \@ifnum [1]{%
 \ifnum #1\expandafter \@firstoftwo
 \else \expandafter \@secondoftwo
 \fi
}%
\providecommand \@ifx [1]{%
 \ifx #1\expandafter \@firstoftwo
 \else \expandafter \@secondoftwo
 \fi
}%
\providecommand \natexlab [1]{#1}%
\providecommand \enquote  [1]{``#1''}%
\providecommand \bibnamefont  [1]{#1}%
\providecommand \bibfnamefont [1]{#1}%
\providecommand \citenamefont [1]{#1}%
\providecommand \href@noop [0]{\@secondoftwo}%
\providecommand \href [0]{\begingroup \@sanitize@url \@href}%
\providecommand \@href[1]{\@@startlink{#1}\@@href}%
\providecommand \@@href[1]{\endgroup#1\@@endlink}%
\providecommand \@sanitize@url [0]{\catcode `\\12\catcode `\$12\catcode
  `\&12\catcode `\#12\catcode `\^12\catcode `\_12\catcode `\%12\relax}%
\providecommand \@@startlink[1]{}%
\providecommand \@@endlink[0]{}%
\providecommand \url  [0]{\begingroup\@sanitize@url \@url }%
\providecommand \@url [1]{\endgroup\@href {#1}{\urlprefix }}%
\providecommand \urlprefix  [0]{URL }%
\providecommand \Eprint [0]{\href }%
\providecommand \doibase [0]{http://dx.doi.org/}%
\providecommand \selectlanguage [0]{\@gobble}%
\providecommand \bibinfo  [0]{\@secondoftwo}%
\providecommand \bibfield  [0]{\@secondoftwo}%
\providecommand \translation [1]{[#1]}%
\providecommand \BibitemOpen [0]{}%
\providecommand \bibitemStop [0]{}%
\providecommand \bibitemNoStop [0]{.\EOS\space}%
\providecommand \EOS [0]{\spacefactor3000\relax}%
\providecommand \BibitemShut  [1]{\csname bibitem#1\endcsname}%
\let\auto@bib@innerbib\@empty
\bibitem [{\citenamefont {Mannhart}\ and\ \citenamefont
  {Schlom}(2010)}]{Mannhart/Schlom:2010}%
  \BibitemOpen
  \bibfield  {author} {\bibinfo {author} {\bibfnamefont {J.}~\bibnamefont
  {Mannhart}}\ and\ \bibinfo {author} {\bibfnamefont {D.~G.}\ \bibnamefont
  {Schlom}},\ }\href {\doibase 10.1126/science.1181862} {\bibfield  {journal}
  {\bibinfo  {journal} {Science}\ }\textbf {\bibinfo {volume} {327}},\ \bibinfo
  {pages} {1607} (\bibinfo {year} {2010})}\BibitemShut {NoStop}%
\bibitem [{\citenamefont {Hwang}\ \emph {et~al.}(2012)\citenamefont {Hwang},
  \citenamefont {Iwasa}, \citenamefont {Kawasaki}, \citenamefont {Keimer},
  \citenamefont {Nagaosa},\ and\ \citenamefont {Tokura}}]{Hwang_et_al:2012}%
  \BibitemOpen
  \bibfield  {author} {\bibinfo {author} {\bibfnamefont {H.~Y.}\ \bibnamefont
  {Hwang}}, \bibinfo {author} {\bibfnamefont {Y.}~\bibnamefont {Iwasa}},
  \bibinfo {author} {\bibfnamefont {M.}~\bibnamefont {Kawasaki}}, \bibinfo
  {author} {\bibfnamefont {B.}~\bibnamefont {Keimer}}, \bibinfo {author}
  {\bibfnamefont {N.}~\bibnamefont {Nagaosa}}, \ and\ \bibinfo {author}
  {\bibfnamefont {Y.}~\bibnamefont {Tokura}},\ }\href {\doibase
  10.1038/nmat3223} {\bibfield  {journal} {\bibinfo  {journal} {Nature
  Materials}\ }\textbf {\bibinfo {volume} {11}},\ \bibinfo {pages} {103}
  (\bibinfo {year} {2012})}\BibitemShut {NoStop}%
\bibitem [{\citenamefont {Ohtomo}\ and\ \citenamefont
  {Hwang}(2004)}]{ohtomo2004high}%
  \BibitemOpen
  \bibfield  {author} {\bibinfo {author} {\bibfnamefont {A.}~\bibnamefont
  {Ohtomo}}\ and\ \bibinfo {author} {\bibfnamefont {H.}~\bibnamefont {Hwang}},\
  }\href@noop {} {\bibfield  {journal} {\bibinfo  {journal} {Nature}\ }\textbf
  {\bibinfo {volume} {427}},\ \bibinfo {pages} {423} (\bibinfo {year}
  {2004})}\BibitemShut {NoStop}%
\bibitem [{\citenamefont {Thiel}\ \emph {et~al.}(2006)\citenamefont {Thiel},
  \citenamefont {Hammerl}, \citenamefont {Schmehl}, \citenamefont {Schneider},\
  and\ \citenamefont {Mannhart}}]{Thiel_et_al:2006}%
  \BibitemOpen
  \bibfield  {author} {\bibinfo {author} {\bibfnamefont {S.}~\bibnamefont
  {Thiel}}, \bibinfo {author} {\bibfnamefont {G.}~\bibnamefont {Hammerl}},
  \bibinfo {author} {\bibfnamefont {A.}~\bibnamefont {Schmehl}}, \bibinfo
  {author} {\bibfnamefont {C.~W.}\ \bibnamefont {Schneider}}, \ and\ \bibinfo
  {author} {\bibfnamefont {J.}~\bibnamefont {Mannhart}},\ }\href@noop {}
  {\bibfield  {journal} {\bibinfo  {journal} {Science}\ }\textbf {\bibinfo
  {volume} {313}},\ \bibinfo {pages} {1942} (\bibinfo {year}
  {2006})}\BibitemShut {NoStop}%
\bibitem [{\citenamefont {Brinkman}\ \emph {et~al.}(2007)\citenamefont
  {Brinkman}, \citenamefont {Huijben}, \citenamefont {Zalk}, \citenamefont
  {Huijben}, \citenamefont {Zeitler}, \citenamefont {Maan}, \citenamefont
  {Wiel}, \citenamefont {Rijnders}, \citenamefont {Blank},\ and\ \citenamefont
  {Hilgenkamp}}]{Brinkman_et_al:2007}%
  \BibitemOpen
  \bibfield  {author} {\bibinfo {author} {\bibfnamefont {A.}~\bibnamefont
  {Brinkman}}, \bibinfo {author} {\bibfnamefont {M.}~\bibnamefont {Huijben}},
  \bibinfo {author} {\bibfnamefont {M.~V.}\ \bibnamefont {Zalk}}, \bibinfo
  {author} {\bibfnamefont {J.}~\bibnamefont {Huijben}}, \bibinfo {author}
  {\bibfnamefont {U.}~\bibnamefont {Zeitler}}, \bibinfo {author} {\bibfnamefont
  {J.~C.}\ \bibnamefont {Maan}}, \bibinfo {author} {\bibfnamefont {W.~G.
  V.~d.}\ \bibnamefont {Wiel}}, \bibinfo {author} {\bibfnamefont
  {G.}~\bibnamefont {Rijnders}}, \bibinfo {author} {\bibfnamefont {D.~H.~A.}\
  \bibnamefont {Blank}}, \ and\ \bibinfo {author} {\bibfnamefont
  {H.}~\bibnamefont {Hilgenkamp}},\ }\href@noop {} {\bibfield  {journal}
  {\bibinfo  {journal} {Nature Materials}\ }\textbf {\bibinfo {volume} {6}},\
  \bibinfo {pages} {493} (\bibinfo {year} {2007})}\BibitemShut {NoStop}%
\bibitem [{\citenamefont {Reyren}\ \emph {et~al.}(2007)\citenamefont {Reyren},
  \citenamefont {Thiel}, \citenamefont {Caviglia}, \citenamefont {Kourkoutis},
  \citenamefont {Hammerl}, \citenamefont {Richter}, \citenamefont {Schneider},
  \citenamefont {Kopp}, \citenamefont {R{\"{u}}etschi}, \citenamefont
  {Jaccard}, \citenamefont {Gabay}, \citenamefont {Muller}, \citenamefont
  {Triscone},\ and\ \citenamefont {Mannhart}}]{Reyren_et_al:2007}%
  \BibitemOpen
  \bibfield  {author} {\bibinfo {author} {\bibfnamefont {N.}~\bibnamefont
  {Reyren}}, \bibinfo {author} {\bibfnamefont {S.}~\bibnamefont {Thiel}},
  \bibinfo {author} {\bibfnamefont {A.~D.}\ \bibnamefont {Caviglia}}, \bibinfo
  {author} {\bibfnamefont {L.~F.}\ \bibnamefont {Kourkoutis}}, \bibinfo
  {author} {\bibfnamefont {G.}~\bibnamefont {Hammerl}}, \bibinfo {author}
  {\bibfnamefont {C.}~\bibnamefont {Richter}}, \bibinfo {author} {\bibfnamefont
  {C.~W.}\ \bibnamefont {Schneider}}, \bibinfo {author} {\bibfnamefont
  {T.}~\bibnamefont {Kopp}}, \bibinfo {author} {\bibfnamefont {A.~S.}\
  \bibnamefont {R{\"{u}}etschi}}, \bibinfo {author} {\bibfnamefont
  {D.}~\bibnamefont {Jaccard}}, \bibinfo {author} {\bibfnamefont
  {M.}~\bibnamefont {Gabay}}, \bibinfo {author} {\bibfnamefont {D.~A.}\
  \bibnamefont {Muller}}, \bibinfo {author} {\bibfnamefont {J.~M.}\
  \bibnamefont {Triscone}}, \ and\ \bibinfo {author} {\bibfnamefont
  {J.}~\bibnamefont {Mannhart}},\ }\href@noop {} {\bibfield  {journal}
  {\bibinfo  {journal} {Science}\ }\textbf {\bibinfo {volume} {317}},\ \bibinfo
  {pages} {1196} (\bibinfo {year} {2007})}\BibitemShut {NoStop}%
\bibitem [{\citenamefont {Ohtomo}\ \emph {et~al.}(2002)\citenamefont {Ohtomo},
  \citenamefont {Muller}, \citenamefont {Grazul},\ and\ \citenamefont
  {Hwang}}]{ohtomo2002artificial}%
  \BibitemOpen
  \bibfield  {author} {\bibinfo {author} {\bibfnamefont {A.}~\bibnamefont
  {Ohtomo}}, \bibinfo {author} {\bibfnamefont {D.}~\bibnamefont {Muller}},
  \bibinfo {author} {\bibfnamefont {J.}~\bibnamefont {Grazul}}, \ and\ \bibinfo
  {author} {\bibfnamefont {H.~Y.}\ \bibnamefont {Hwang}},\ }\href@noop {}
  {\bibfield  {journal} {\bibinfo  {journal} {Nature}\ }\textbf {\bibinfo
  {volume} {419}},\ \bibinfo {pages} {378} (\bibinfo {year}
  {2002})}\BibitemShut {NoStop}%
\bibitem [{\citenamefont {Georges}\ \emph {et~al.}(1996)\citenamefont
  {Georges}, \citenamefont {Kotliar}, \citenamefont {Krauth},\ and\
  \citenamefont {Rozenberg}}]{Georges_et_al:1996}%
  \BibitemOpen
  \bibfield  {author} {\bibinfo {author} {\bibfnamefont {A.}~\bibnamefont
  {Georges}}, \bibinfo {author} {\bibfnamefont {G.}~\bibnamefont {Kotliar}},
  \bibinfo {author} {\bibfnamefont {W.}~\bibnamefont {Krauth}}, \ and\ \bibinfo
  {author} {\bibfnamefont {M.~J.}\ \bibnamefont {Rozenberg}},\ }\href {\doibase
  10.1103/RevModPhys.68.13} {\bibfield  {journal} {\bibinfo  {journal} {Reviews
  of Modern Physics}\ }\textbf {\bibinfo {volume} {68}},\ \bibinfo {pages} {13}
  (\bibinfo {year} {1996})}\BibitemShut {NoStop}%
\bibitem [{\citenamefont {Anisimov}\ \emph {et~al.}(1997)\citenamefont
  {Anisimov}, \citenamefont {Poteryaev}, \citenamefont {Korotin}, \citenamefont
  {Anokhin},\ and\ \citenamefont {Kotliar}}]{Anisimov_et_al:1997}%
  \BibitemOpen
  \bibfield  {author} {\bibinfo {author} {\bibfnamefont {V.~I.}\ \bibnamefont
  {Anisimov}}, \bibinfo {author} {\bibfnamefont {A.~I.}\ \bibnamefont
  {Poteryaev}}, \bibinfo {author} {\bibfnamefont {M.~A.}\ \bibnamefont
  {Korotin}}, \bibinfo {author} {\bibfnamefont {A.~O.}\ \bibnamefont
  {Anokhin}}, \ and\ \bibinfo {author} {\bibfnamefont {G.}~\bibnamefont
  {Kotliar}},\ }\href@noop {} {\bibfield  {journal} {\bibinfo  {journal}
  {Journal of Physics: Condensed Matter}\ }\textbf {\bibinfo {volume} {9}},\
  \bibinfo {pages} {7359} (\bibinfo {year} {1997})}\BibitemShut {NoStop}%
\bibitem [{\citenamefont {Held}(2007)}]{Held:2007}%
  \BibitemOpen
  \bibfield  {author} {\bibinfo {author} {\bibfnamefont {K.}~\bibnamefont
  {Held}},\ }\href {\doibase 10.1080/00018730701619647} {\bibfield  {journal}
  {\bibinfo  {journal} {Advances in Physics}\ }\textbf {\bibinfo {volume}
  {56}},\ \bibinfo {pages} {829} (\bibinfo {year} {2007})}\BibitemShut
  {NoStop}%
\bibitem [{\citenamefont {Zhong}\ and\ \citenamefont
  {Hansmann}(2017)}]{zhong2017band}%
  \BibitemOpen
  \bibfield  {author} {\bibinfo {author} {\bibfnamefont {Z.}~\bibnamefont
  {Zhong}}\ and\ \bibinfo {author} {\bibfnamefont {P.}~\bibnamefont
  {Hansmann}},\ }\href@noop {} {\bibfield  {journal} {\bibinfo  {journal}
  {Physical Review X}\ }\textbf {\bibinfo {volume} {7}},\ \bibinfo {pages}
  {011023} (\bibinfo {year} {2017})}\BibitemShut {NoStop}%
\bibitem [{\citenamefont {Chen}\ and\ \citenamefont
  {Millis}(2017)}]{Chen/Millis:2017}%
  \BibitemOpen
  \bibfield  {author} {\bibinfo {author} {\bibfnamefont {H.}~\bibnamefont
  {Chen}}\ and\ \bibinfo {author} {\bibfnamefont {A.}~\bibnamefont {Millis}},\
  }\href {\doibase 10.1088/1361-648x/aa6efe} {\bibfield  {journal} {\bibinfo
  {journal} {Journal of Physics: Condensed Matter}\ }\textbf {\bibinfo {volume}
  {29}},\ \bibinfo {pages} {243001} (\bibinfo {year} {2017})}\BibitemShut
  {NoStop}%
\bibitem [{\citenamefont {Giannozzi}\ \emph {et~al.}(2009)\citenamefont
  {Giannozzi}, \citenamefont {Baroni}, \citenamefont {Bonini}, \citenamefont
  {Calandra}, \citenamefont {Car}, \citenamefont {Cavazzoni}, \citenamefont
  {Ceresoli}, \citenamefont {Chiarotti}, \citenamefont {Cococcioni},
  \citenamefont {Dabo}, \citenamefont {Corso}, \citenamefont {de~Gironcoli},
  \citenamefont {Fabris}, \citenamefont {Fratesi}, \citenamefont {Gebauer},
  \citenamefont {Gerstmann}, \citenamefont {Gougoussis}, \citenamefont
  {Kokalj}, \citenamefont {Lazzeri}, \citenamefont {Martin-Samos},
  \citenamefont {Marzari}, \citenamefont {Mauri}, \citenamefont {Mazzarello},
  \citenamefont {Paolini}, \citenamefont {Pasquarello}, \citenamefont
  {Paulatto}, \citenamefont {Sbraccia}, \citenamefont {Scandalo}, \citenamefont
  {Sclauzero}, \citenamefont {Seitsonen}, \citenamefont {Smogunov},
  \citenamefont {Umari},\ and\ \citenamefont
  {Wentzcovitch}}]{Giannozzi_et_al:2009}%
  \BibitemOpen
  \bibfield  {author} {\bibinfo {author} {\bibfnamefont {P.}~\bibnamefont
  {Giannozzi}}, \bibinfo {author} {\bibfnamefont {S.}~\bibnamefont {Baroni}},
  \bibinfo {author} {\bibfnamefont {N.}~\bibnamefont {Bonini}}, \bibinfo
  {author} {\bibfnamefont {M.}~\bibnamefont {Calandra}}, \bibinfo {author}
  {\bibfnamefont {R.}~\bibnamefont {Car}}, \bibinfo {author} {\bibfnamefont
  {C.}~\bibnamefont {Cavazzoni}}, \bibinfo {author} {\bibfnamefont
  {D.}~\bibnamefont {Ceresoli}}, \bibinfo {author} {\bibfnamefont {G.~L.}\
  \bibnamefont {Chiarotti}}, \bibinfo {author} {\bibfnamefont {M.}~\bibnamefont
  {Cococcioni}}, \bibinfo {author} {\bibfnamefont {I.}~\bibnamefont {Dabo}},
  \bibinfo {author} {\bibfnamefont {A.~D.}\ \bibnamefont {Corso}}, \bibinfo
  {author} {\bibfnamefont {S.}~\bibnamefont {de~Gironcoli}}, \bibinfo {author}
  {\bibfnamefont {S.}~\bibnamefont {Fabris}}, \bibinfo {author} {\bibfnamefont
  {G.}~\bibnamefont {Fratesi}}, \bibinfo {author} {\bibfnamefont
  {R.}~\bibnamefont {Gebauer}}, \bibinfo {author} {\bibfnamefont
  {U.}~\bibnamefont {Gerstmann}}, \bibinfo {author} {\bibfnamefont
  {C.}~\bibnamefont {Gougoussis}}, \bibinfo {author} {\bibfnamefont
  {A.}~\bibnamefont {Kokalj}}, \bibinfo {author} {\bibfnamefont
  {M.}~\bibnamefont {Lazzeri}}, \bibinfo {author} {\bibfnamefont
  {L.}~\bibnamefont {Martin-Samos}}, \bibinfo {author} {\bibfnamefont
  {N.}~\bibnamefont {Marzari}}, \bibinfo {author} {\bibfnamefont
  {F.}~\bibnamefont {Mauri}}, \bibinfo {author} {\bibfnamefont
  {R.}~\bibnamefont {Mazzarello}}, \bibinfo {author} {\bibfnamefont
  {S.}~\bibnamefont {Paolini}}, \bibinfo {author} {\bibfnamefont
  {A.}~\bibnamefont {Pasquarello}}, \bibinfo {author} {\bibfnamefont
  {L.}~\bibnamefont {Paulatto}}, \bibinfo {author} {\bibfnamefont
  {C.}~\bibnamefont {Sbraccia}}, \bibinfo {author} {\bibfnamefont
  {S.}~\bibnamefont {Scandalo}}, \bibinfo {author} {\bibfnamefont
  {G.}~\bibnamefont {Sclauzero}}, \bibinfo {author} {\bibfnamefont {A.~P.}\
  \bibnamefont {Seitsonen}}, \bibinfo {author} {\bibfnamefont {A.}~\bibnamefont
  {Smogunov}}, \bibinfo {author} {\bibfnamefont {P.}~\bibnamefont {Umari}}, \
  and\ \bibinfo {author} {\bibfnamefont {R.}~\bibnamefont {Wentzcovitch}},\
  }\href {\doibase 10.1088/0953-8984/21/39/395502} {\bibfield  {journal}
  {\bibinfo  {journal} {Journal of Physics: Condensed Matter}\ }\textbf
  {\bibinfo {volume} {21}},\ \bibinfo {pages} {395502} (\bibinfo {year}
  {2009})}\BibitemShut {NoStop}%
\bibitem [{\citenamefont {Perdew}\ \emph {et~al.}(1996)\citenamefont {Perdew},
  \citenamefont {Burke},\ and\ \citenamefont
  {Ernzerhof}}]{Perdew/Burke/Ernzerhof:1996}%
  \BibitemOpen
  \bibfield  {author} {\bibinfo {author} {\bibfnamefont {J.~P.}\ \bibnamefont
  {Perdew}}, \bibinfo {author} {\bibfnamefont {K.}~\bibnamefont {Burke}}, \
  and\ \bibinfo {author} {\bibfnamefont {M.}~\bibnamefont {Ernzerhof}},\ }\href
  {\doibase 10.1103/PhysRevLett.77.3865} {\bibfield  {journal} {\bibinfo
  {journal} {Physical Review Letters}\ }\textbf {\bibinfo {volume} {77}},\
  \bibinfo {pages} {3865} (\bibinfo {year} {1996})}\BibitemShut {NoStop}%
\bibitem [{\citenamefont {Mostofi}\ \emph {et~al.}(2008)\citenamefont
  {Mostofi}, \citenamefont {Yates}, \citenamefont {Lee}, \citenamefont {Souza},
  \citenamefont {Vanderbilt},\ and\ \citenamefont
  {Marzari}}]{Mostofi_et_al:2008}%
  \BibitemOpen
  \bibfield  {author} {\bibinfo {author} {\bibfnamefont {A.~A.}\ \bibnamefont
  {Mostofi}}, \bibinfo {author} {\bibfnamefont {J.~R.}\ \bibnamefont {Yates}},
  \bibinfo {author} {\bibfnamefont {Y.-S.}\ \bibnamefont {Lee}}, \bibinfo
  {author} {\bibfnamefont {I.}~\bibnamefont {Souza}}, \bibinfo {author}
  {\bibfnamefont {D.}~\bibnamefont {Vanderbilt}}, \ and\ \bibinfo {author}
  {\bibfnamefont {N.}~\bibnamefont {Marzari}},\ }\href {\doibase
  10.1016/j.cpc.2007.11.016} {\bibfield  {journal} {\bibinfo  {journal}
  {Computer Physics Communications}\ }\textbf {\bibinfo {volume} {178}},\
  \bibinfo {pages} {685} (\bibinfo {year} {2008})}\BibitemShut {NoStop}%
\bibitem [{\citenamefont {Marzari}\ \emph {et~al.}(2012)\citenamefont
  {Marzari}, \citenamefont {Mostofi}, \citenamefont {Yates}, \citenamefont
  {Souza},\ and\ \citenamefont {Vanderbilt}}]{Marzari_et_al:2012}%
  \BibitemOpen
  \bibfield  {author} {\bibinfo {author} {\bibfnamefont {N.}~\bibnamefont
  {Marzari}}, \bibinfo {author} {\bibfnamefont {A.~A.}\ \bibnamefont
  {Mostofi}}, \bibinfo {author} {\bibfnamefont {J.~R.}\ \bibnamefont {Yates}},
  \bibinfo {author} {\bibfnamefont {I.}~\bibnamefont {Souza}}, \ and\ \bibinfo
  {author} {\bibfnamefont {D.}~\bibnamefont {Vanderbilt}},\ }\href {\doibase
  10.1103/RevModPhys.84.1419} {\bibfield  {journal} {\bibinfo  {journal}
  {Reviews of Modern Physics}\ }\textbf {\bibinfo {volume} {84}},\ \bibinfo
  {pages} {1419} (\bibinfo {year} {2012})}\BibitemShut {NoStop}%
\bibitem [{\citenamefont {Seth}\ \emph {et~al.}(2016)\citenamefont {Seth},
  \citenamefont {Krivenko}, \citenamefont {Ferrero},\ and\ \citenamefont
  {Parcollet}}]{Seth2016274}%
  \BibitemOpen
  \bibfield  {author} {\bibinfo {author} {\bibfnamefont {P.}~\bibnamefont
  {Seth}}, \bibinfo {author} {\bibfnamefont {I.}~\bibnamefont {Krivenko}},
  \bibinfo {author} {\bibfnamefont {M.}~\bibnamefont {Ferrero}}, \ and\
  \bibinfo {author} {\bibfnamefont {O.}~\bibnamefont {Parcollet}},\ }\href
  {\doibase http://dx.doi.org/10.1016/j.cpc.2015.10.023} {\bibfield  {journal}
  {\bibinfo  {journal} {Computer Physics Communications}\ }\textbf {\bibinfo
  {volume} {200}},\ \bibinfo {pages} {274 } (\bibinfo {year}
  {2016})}\BibitemShut {NoStop}%
\bibitem [{\citenamefont {Sclauzero}\ \emph {et~al.}(2016)\citenamefont
  {Sclauzero}, \citenamefont {Dymkowski},\ and\ \citenamefont
  {Ederer}}]{Sclauzero/Dymkowski/Ederer:2016}%
  \BibitemOpen
  \bibfield  {author} {\bibinfo {author} {\bibfnamefont {G.}~\bibnamefont
  {Sclauzero}}, \bibinfo {author} {\bibfnamefont {K.}~\bibnamefont
  {Dymkowski}}, \ and\ \bibinfo {author} {\bibfnamefont {C.}~\bibnamefont
  {Ederer}},\ }\href {\doibase 10.1103/PhysRevB.94.245109} {\bibfield
  {journal} {\bibinfo  {journal} {Physical Review B}\ }\textbf {\bibinfo
  {volume} {94}},\ \bibinfo {pages} {245109} (\bibinfo {year}
  {2016})}\BibitemShut {NoStop}%
\bibitem [{\citenamefont {Pavarini}\ \emph {et~al.}(2004)\citenamefont
  {Pavarini}, \citenamefont {Biermann}, \citenamefont {Poteryaev},
  \citenamefont {Lichtenstein}, \citenamefont {Georges},\ and\ \citenamefont
  {Andersen}}]{Pavarini_et_al:2004}%
  \BibitemOpen
  \bibfield  {author} {\bibinfo {author} {\bibfnamefont {E.}~\bibnamefont
  {Pavarini}}, \bibinfo {author} {\bibfnamefont {S.}~\bibnamefont {Biermann}},
  \bibinfo {author} {\bibfnamefont {A.}~\bibnamefont {Poteryaev}}, \bibinfo
  {author} {\bibfnamefont {A.~I.}\ \bibnamefont {Lichtenstein}}, \bibinfo
  {author} {\bibfnamefont {A.}~\bibnamefont {Georges}}, \ and\ \bibinfo
  {author} {\bibfnamefont {O.~K.}\ \bibnamefont {Andersen}},\ }\href {\doibase
  10.1103/PhysRevLett.92.176403} {\bibfield  {journal} {\bibinfo  {journal}
  {Physical Review Letters}\ }\textbf {\bibinfo {volume} {92}},\ \bibinfo
  {pages} {176403} (\bibinfo {year} {2004})}\BibitemShut {NoStop}%
\bibitem [{\citenamefont {De~Raychaudhury}\ \emph {et~al.}(2007)\citenamefont
  {De~Raychaudhury}, \citenamefont {Pavarini},\ and\ \citenamefont
  {Andersen}}]{DeRaychaudhury/Pavarini/Andersen:2007}%
  \BibitemOpen
  \bibfield  {author} {\bibinfo {author} {\bibfnamefont {M.}~\bibnamefont
  {De~Raychaudhury}}, \bibinfo {author} {\bibfnamefont {E.}~\bibnamefont
  {Pavarini}}, \ and\ \bibinfo {author} {\bibfnamefont {O.~K.}\ \bibnamefont
  {Andersen}},\ }\href {http://dx.doi.org/10.1103/PhysRevLett.99.126402}
  {\bibfield  {journal} {\bibinfo  {journal} {Physical Review Letters}\
  }\textbf {\bibinfo {volume} {99}},\ \bibinfo {pages} {126402} (\bibinfo
  {year} {2007})}\BibitemShut {NoStop}%
\bibitem [{\citenamefont {Dymkowski}\ and\ \citenamefont
  {Ederer}(2014)}]{Dymkowski/Ederer:2014}%
  \BibitemOpen
  \bibfield  {author} {\bibinfo {author} {\bibfnamefont {K.}~\bibnamefont
  {Dymkowski}}\ and\ \bibinfo {author} {\bibfnamefont {C.}~\bibnamefont
  {Ederer}},\ }\href {\doibase 10.1103/PhysRevB.89.161109} {\bibfield
  {journal} {\bibinfo  {journal} {Physical Review B}\ }\textbf {\bibinfo
  {volume} {89}},\ \bibinfo {pages} {161109} (\bibinfo {year}
  {2014})}\BibitemShut {NoStop}%
\bibitem [{\citenamefont {Sclauzero}\ and\ \citenamefont
  {Ederer}(2015)}]{Sclauzero/Ederer:2015}%
  \BibitemOpen
  \bibfield  {author} {\bibinfo {author} {\bibfnamefont {G.}~\bibnamefont
  {Sclauzero}}\ and\ \bibinfo {author} {\bibfnamefont {C.}~\bibnamefont
  {Ederer}},\ }\href {http://dx.doi.org/10.1103/PhysRevB.92.235112} {\bibfield
  {journal} {\bibinfo  {journal} {Physical Review B}\ }\textbf {\bibinfo
  {volume} {92}},\ \bibinfo {pages} {235112} (\bibinfo {year}
  {2015})}\BibitemShut {NoStop}%
\bibitem [{\citenamefont {Aichhorn}\ \emph {et~al.}(2016)\citenamefont
  {Aichhorn}, \citenamefont {Pourovskii}, \citenamefont {Seth}, \citenamefont
  {Vildosola}, \citenamefont {Zingl}, \citenamefont {Peil}, \citenamefont
  {Deng}, \citenamefont {Mravlje}, \citenamefont {Kraberger}, \citenamefont
  {Martins}, \citenamefont {Ferrero},\ and\ \citenamefont
  {Parcollet}}]{aichhorn_dfttools_2016}%
  \BibitemOpen
  \bibfield  {author} {\bibinfo {author} {\bibfnamefont {M.}~\bibnamefont
  {Aichhorn}}, \bibinfo {author} {\bibfnamefont {L.}~\bibnamefont
  {Pourovskii}}, \bibinfo {author} {\bibfnamefont {P.}~\bibnamefont {Seth}},
  \bibinfo {author} {\bibfnamefont {V.}~\bibnamefont {Vildosola}}, \bibinfo
  {author} {\bibfnamefont {M.}~\bibnamefont {Zingl}}, \bibinfo {author}
  {\bibfnamefont {O.}~\bibnamefont {Peil}}, \bibinfo {author} {\bibfnamefont
  {X.}~\bibnamefont {Deng}}, \bibinfo {author} {\bibfnamefont {J.}~\bibnamefont
  {Mravlje}}, \bibinfo {author} {\bibfnamefont {G.}~\bibnamefont {Kraberger}},
  \bibinfo {author} {\bibfnamefont {C.}~\bibnamefont {Martins}}, \bibinfo
  {author} {\bibfnamefont {M.}~\bibnamefont {Ferrero}}, \ and\ \bibinfo
  {author} {\bibfnamefont {O.}~\bibnamefont {Parcollet}},\ }\href {\doibase
  doi:10.1016/j.cpc.2016.03.014} {\bibfield  {journal} {\bibinfo  {journal}
  {Computer Physics Communications}\ }\textbf {\bibinfo {volume} {204}},\
  \bibinfo {pages} {200} (\bibinfo {year} {2016})}\BibitemShut {NoStop}%
\bibitem [{\citenamefont {Fuchs}\ \emph {et~al.}(2011)\citenamefont {Fuchs},
  \citenamefont {Gull}, \citenamefont {Troyer}, \citenamefont {Jarrell},\ and\
  \citenamefont {Pruschke}}]{Fuchs_et_al:2011}%
  \BibitemOpen
  \bibfield  {author} {\bibinfo {author} {\bibfnamefont {S.}~\bibnamefont
  {Fuchs}}, \bibinfo {author} {\bibfnamefont {E.}~\bibnamefont {Gull}},
  \bibinfo {author} {\bibfnamefont {M.}~\bibnamefont {Troyer}}, \bibinfo
  {author} {\bibfnamefont {M.}~\bibnamefont {Jarrell}}, \ and\ \bibinfo
  {author} {\bibfnamefont {T.}~\bibnamefont {Pruschke}},\ }\href
  {http://dx.doi.org/10.1103/PhysRevB.83.235113} {\bibfield  {journal}
  {\bibinfo  {journal} {Physical Review B}\ }\textbf {\bibinfo {volume} {83}},\
  \bibinfo {pages} {235113} (\bibinfo {year} {2011})}\BibitemShut {NoStop}%
\bibitem [{\citenamefont {Jarrell}\ and\ \citenamefont
  {Gubernatis}(1996)}]{jarrell1996bayesian}%
  \BibitemOpen
  \bibfield  {author} {\bibinfo {author} {\bibfnamefont {M.}~\bibnamefont
  {Jarrell}}\ and\ \bibinfo {author} {\bibfnamefont {J.~E.}\ \bibnamefont
  {Gubernatis}},\ }\href@noop {} {\bibfield  {journal} {\bibinfo  {journal}
  {Physics Reports}\ }\textbf {\bibinfo {volume} {269}},\ \bibinfo {pages}
  {133} (\bibinfo {year} {1996})}\BibitemShut {NoStop}%
\bibitem [{\citenamefont {Rondinelli}\ and\ \citenamefont
  {Spaldin}(2011)}]{Rondinelli:2011}%
  \BibitemOpen
  \bibfield  {author} {\bibinfo {author} {\bibfnamefont {J.~M.}\ \bibnamefont
  {Rondinelli}}\ and\ \bibinfo {author} {\bibfnamefont {N.~A.}\ \bibnamefont
  {Spaldin}},\ }\href {\doibase 10.1002/adma.201101152} {\bibfield  {journal}
  {\bibinfo  {journal} {Advanced Materials}\ }\textbf {\bibinfo {volume}
  {23}},\ \bibinfo {pages} {3363} (\bibinfo {year} {2011})}\BibitemShut
  {NoStop}%
\bibitem [{\citenamefont {Park}\ \emph {et~al.}(2017)\citenamefont {Park},
  \citenamefont {Kumar},\ and\ \citenamefont {Rabe}}]{park2017charge}%
  \BibitemOpen
  \bibfield  {author} {\bibinfo {author} {\bibfnamefont {S.~Y.}\ \bibnamefont
  {Park}}, \bibinfo {author} {\bibfnamefont {A.}~\bibnamefont {Kumar}}, \ and\
  \bibinfo {author} {\bibfnamefont {K.~M.}\ \bibnamefont {Rabe}},\ }\href@noop
  {} {\bibfield  {journal} {\bibinfo  {journal} {Physical Review Letters}\
  }\textbf {\bibinfo {volume} {118}},\ \bibinfo {pages} {087602} (\bibinfo
  {year} {2017})}\BibitemShut {NoStop}%
\bibitem [{\citenamefont {Tan}\ \emph {et~al.}(2013)\citenamefont {Tan},
  \citenamefont {Egoavil}, \citenamefont {B\'ech\'e}, \citenamefont {Martinez},
  \citenamefont {Van~Aert}, \citenamefont {Verbeeck}, \citenamefont
  {Van~Tendeloo}, \citenamefont {Rotella}, \citenamefont {Boullay},
  \citenamefont {Pautrat},\ and\ \citenamefont {Prellier}}]{tan2013mapping}%
  \BibitemOpen
  \bibfield  {author} {\bibinfo {author} {\bibfnamefont {H.}~\bibnamefont
  {Tan}}, \bibinfo {author} {\bibfnamefont {R.}~\bibnamefont {Egoavil}},
  \bibinfo {author} {\bibfnamefont {A.}~\bibnamefont {B\'ech\'e}}, \bibinfo
  {author} {\bibfnamefont {G.~T.}\ \bibnamefont {Martinez}}, \bibinfo {author}
  {\bibfnamefont {S.}~\bibnamefont {Van~Aert}}, \bibinfo {author}
  {\bibfnamefont {J.}~\bibnamefont {Verbeeck}}, \bibinfo {author}
  {\bibfnamefont {G.}~\bibnamefont {Van~Tendeloo}}, \bibinfo {author}
  {\bibfnamefont {H.}~\bibnamefont {Rotella}}, \bibinfo {author} {\bibfnamefont
  {P.}~\bibnamefont {Boullay}}, \bibinfo {author} {\bibfnamefont
  {A.}~\bibnamefont {Pautrat}}, \ and\ \bibinfo {author} {\bibfnamefont
  {W.}~\bibnamefont {Prellier}},\ }\href {\doibase 10.1103/PhysRevB.88.155123}
  {\bibfield  {journal} {\bibinfo  {journal} {Physical Review B}\ }\textbf
  {\bibinfo {volume} {88}},\ \bibinfo {pages} {155123} (\bibinfo {year}
  {2013})}\BibitemShut {NoStop}%
\bibitem [{\citenamefont {Weng}\ \emph {et~al.}(2017)\citenamefont {Weng},
  \citenamefont {Zhang}, \citenamefont {Gao},\ and\ \citenamefont
  {Dong}}]{weng2017latio}%
  \BibitemOpen
  \bibfield  {author} {\bibinfo {author} {\bibfnamefont {Y.}~\bibnamefont
  {Weng}}, \bibinfo {author} {\bibfnamefont {J.-J.}\ \bibnamefont {Zhang}},
  \bibinfo {author} {\bibfnamefont {B.}~\bibnamefont {Gao}}, \ and\ \bibinfo
  {author} {\bibfnamefont {S.}~\bibnamefont {Dong}},\ }\href@noop {} {\bibfield
   {journal} {\bibinfo  {journal} {Physical Review B}\ }\textbf {\bibinfo
  {volume} {95}},\ \bibinfo {pages} {155117} (\bibinfo {year}
  {2017})}\BibitemShut {NoStop}%
\bibitem [{\citenamefont {He}\ \emph {et~al.}(2012)\citenamefont {He},
  \citenamefont {Sanders}, \citenamefont {Gray}, \citenamefont {Wong},
  \citenamefont {Mehta},\ and\ \citenamefont {Suzuki}}]{He_et_al:2012}%
  \BibitemOpen
  \bibfield  {author} {\bibinfo {author} {\bibfnamefont {C.}~\bibnamefont
  {He}}, \bibinfo {author} {\bibfnamefont {T.~D.}\ \bibnamefont {Sanders}},
  \bibinfo {author} {\bibfnamefont {M.~T.}\ \bibnamefont {Gray}}, \bibinfo
  {author} {\bibfnamefont {F.~J.}\ \bibnamefont {Wong}}, \bibinfo {author}
  {\bibfnamefont {V.~V.}\ \bibnamefont {Mehta}}, \ and\ \bibinfo {author}
  {\bibfnamefont {Y.}~\bibnamefont {Suzuki}},\ }\href {\doibase
  10.1103/PhysRevB.86.081401} {\bibfield  {journal} {\bibinfo  {journal}
  {Physical Review B}\ }\textbf {\bibinfo {volume} {86}},\ \bibinfo {pages}
  {081401} (\bibinfo {year} {2012})}\BibitemShut {NoStop}%
\bibitem [{\citenamefont {L\"uders}\ \emph {et~al.}(2014)\citenamefont
  {L\"uders}, \citenamefont {Li}, \citenamefont {Feyerherm},\ and\
  \citenamefont {Dudzik}}]{Luders_et_al:2014}%
  \BibitemOpen
  \bibfield  {author} {\bibinfo {author} {\bibfnamefont {U.}~\bibnamefont
  {L\"uders}}, \bibinfo {author} {\bibfnamefont {Q.-R.}\ \bibnamefont {Li}},
  \bibinfo {author} {\bibfnamefont {R.}~\bibnamefont {Feyerherm}}, \ and\
  \bibinfo {author} {\bibfnamefont {E.}~\bibnamefont {Dudzik}},\ }\href
  {\doibase https://doi.org/10.1016/j.jpcs.2014.07.007} {\bibfield  {journal}
  {\bibinfo  {journal} {Journal of Physics and Chemistry of Solids}\ }\textbf
  {\bibinfo {volume} {75}},\ \bibinfo {pages} {1354 } (\bibinfo {year}
  {2014})}\BibitemShut {NoStop}%
\bibitem [{\citenamefont {Eylem}\ \emph {et~al.}(1996)\citenamefont {Eylem},
  \citenamefont {Hung}, \citenamefont {Ju}, \citenamefont {Kim}, \citenamefont
  {Green}, \citenamefont {Vogt}, \citenamefont {Hriljac}, \citenamefont
  {Eichhorn}, \citenamefont {Greene},\ and\ \citenamefont
  {Salamanca-Riba}}]{eylem1996unusual}%
  \BibitemOpen
  \bibfield  {author} {\bibinfo {author} {\bibfnamefont {C.}~\bibnamefont
  {Eylem}}, \bibinfo {author} {\bibfnamefont {Y.-C.}\ \bibnamefont {Hung}},
  \bibinfo {author} {\bibfnamefont {H.}~\bibnamefont {Ju}}, \bibinfo {author}
  {\bibfnamefont {J.}~\bibnamefont {Kim}}, \bibinfo {author} {\bibfnamefont
  {D.}~\bibnamefont {Green}}, \bibinfo {author} {\bibfnamefont
  {T.}~\bibnamefont {Vogt}}, \bibinfo {author} {\bibfnamefont {J.}~\bibnamefont
  {Hriljac}}, \bibinfo {author} {\bibfnamefont {B.}~\bibnamefont {Eichhorn}},
  \bibinfo {author} {\bibfnamefont {R.}~\bibnamefont {Greene}}, \ and\ \bibinfo
  {author} {\bibfnamefont {L.}~\bibnamefont {Salamanca-Riba}},\ }\href@noop {}
  {\bibfield  {journal} {\bibinfo  {journal} {Chemistry of Materials}\ }\textbf
  {\bibinfo {volume} {8}},\ \bibinfo {pages} {418} (\bibinfo {year}
  {1996})}\BibitemShut {NoStop}%
\bibitem [{\citenamefont {Chen}\ \emph {et~al.}(2013)\citenamefont {Chen},
  \citenamefont {Millis},\ and\ \citenamefont {Marianetti}}]{Chen_et_al:2013}%
  \BibitemOpen
  \bibfield  {author} {\bibinfo {author} {\bibfnamefont {H.}~\bibnamefont
  {Chen}}, \bibinfo {author} {\bibfnamefont {A.~J.}\ \bibnamefont {Millis}}, \
  and\ \bibinfo {author} {\bibfnamefont {C.~A.}\ \bibnamefont {Marianetti}},\
  }\href {\doibase 10.1103/PhysRevLett.111.116403} {\bibfield  {journal}
  {\bibinfo  {journal} {Physical Review Letters}\ }\textbf {\bibinfo {volume}
  {111}},\ \bibinfo {pages} {116403} (\bibinfo {year} {2013})}\BibitemShut
  {NoStop}%
\bibitem [{\citenamefont {Kleibeuker}\ \emph {et~al.}(2014)\citenamefont
  {Kleibeuker}, \citenamefont {Zhong}, \citenamefont {Nishikawa}, \citenamefont
  {Gabel}, \citenamefont {M\"uller}, \citenamefont {Pfaff}, \citenamefont
  {Sing}, \citenamefont {Held}, \citenamefont {Claessen}, \citenamefont
  {Koster},\ and\ \citenamefont {Rijnders}}]{Kleibeuker_et_al:2014}%
  \BibitemOpen
  \bibfield  {author} {\bibinfo {author} {\bibfnamefont {J.~E.}\ \bibnamefont
  {Kleibeuker}}, \bibinfo {author} {\bibfnamefont {Z.}~\bibnamefont {Zhong}},
  \bibinfo {author} {\bibfnamefont {H.}~\bibnamefont {Nishikawa}}, \bibinfo
  {author} {\bibfnamefont {J.}~\bibnamefont {Gabel}}, \bibinfo {author}
  {\bibfnamefont {A.}~\bibnamefont {M\"uller}}, \bibinfo {author}
  {\bibfnamefont {F.}~\bibnamefont {Pfaff}}, \bibinfo {author} {\bibfnamefont
  {M.}~\bibnamefont {Sing}}, \bibinfo {author} {\bibfnamefont {K.}~\bibnamefont
  {Held}}, \bibinfo {author} {\bibfnamefont {R.}~\bibnamefont {Claessen}},
  \bibinfo {author} {\bibfnamefont {G.}~\bibnamefont {Koster}}, \ and\ \bibinfo
  {author} {\bibfnamefont {G.}~\bibnamefont {Rijnders}},\ }\href {\doibase
  10.1103/PhysRevLett.113.237402} {\bibfield  {journal} {\bibinfo  {journal}
  {Physical Review Letters}\ }\textbf {\bibinfo {volume} {113}},\ \bibinfo
  {pages} {237402} (\bibinfo {year} {2014})}\BibitemShut {NoStop}%
\end{thebibliography}%

\end{document}